# PolyFold: an interactive visual simulator for distance-based protein folding

Andrew J. McGehee[1], Sutanu Bhattacharya[1,¶], Rahmatullah Roche[1,¶], Debswapna Bhattacharya[1,2*]

[1]Department of Computer Science and Software Engineering, Auburn University, Auburn, AL 36849, USA

[2]Department of Biological Sciences, Auburn University, Auburn, AL 36849, USA

*Corresponding author

E-mail: bhattacharyad@auburn.edu (DB)

¶These authors contributed equally to this work.

# Abstract

Recent advances in distance-based protein folding have led to a paradigm shift in protein structure prediction. Through sufficiently precise estimation of the inter-residue distance matrix for a protein sequence, it is now feasible to predict the correct folds for new proteins much more accurately than ever before. Despite the exciting progress, a dedicated visualization system that can dynamically capture the distance-based folding process is still lacking. Most molecular visualizers typically provide only a static view of a folded protein conformation, but do not capture the folding process. Even among the selected few graphical interfaces that do adopt a dynamic perspective, none of them are distance-based. Here we present PolyFold, an interactive visual simulator for dynamically capturing the distance-based protein folding process through real-time rendering of a distance matrix and its compatible spatial conformation as it folds in an intuitive and easy-to-use interface. PolyFold integrates highly convergent stochastic optimization algorithms with on-demand customizations and interactive manipulations to maximally satisfy the geometric constraints imposed by a distance matrix. PolyFold is capable of simulating the complex process of protein folding even on modest personal computers, thus making it accessible to the general public for fostering citizen science. Open source code of PolyFold is freely available for download at https://github.com/Bhattacharya-Lab/PolyFold. It is implemented in cross-platform Java and binary executables are available for macOS, Linux, and Windows.

# Introduction

Computational protein structure prediction has witnessed remarkable progress in the recent past due to advances in folding new proteins from scratch using sufficiently accurate estimation of the inter-residue distance matrix [1–4]. A distance matrix encodes a protein's three-dimensional (3D) structure through inter-residue spatial proximity information that can be converted to physical constraints in order to drive the ab initio folding process with minimal conformational search [5,6]. Consequently, distance-based protein folding has gained a lot of attention, fueling considerable research efforts [7–11]. However, the lack of a dedicated visualization system that can dynamically capture the distance-based folding process precludes the possibility of obtaining a visual understanding of its nature. Currently popular molecular visualization tools like PyMol and UCSF Chimera [12,13] typically provide only a static view of a folded protein conformation, but do not capture the folding process. Recent graphical interfaces such as the PyRosetta Toolkit [14] and InteractiveROSETTA [15] adopt dynamic perspectives, but they are built exclusively for the ROSETTA molecular modeling suite [16], which primarily relies on a fragment-based approach for protein folding. InteractiveROSETTA has many sophisticated ROSETTA-based features, including APIs to incorporate various distance restraints. However, a standalone visualizer that provides insights to distance-based folding for researchers is still lacking. Beyond the realm of expert-oriented visualization tools, the interactive graphical interface Foldit Standalone [17] makes it possible for non-experts to manipulate protein structures in the context of the popular scientific discovery game Foldit [18], which itself is based on ROSETTA and thus not distance-based. A dedicated distance-based visual folding simulator with a simple to use interface will not only provide a central platform for researchers to delve deeper into the folding process and gain critical insights, but will also make the latest technological advances in protein folding and molecular modeling easily accessible to non-experts, while still being scientifically accurate.

We have developed a brand-new standalone GUI called PolyFold for visually simulating the distance-based protein folding process. PolyFold provides several user-friendly controls for running powerful distance matrix optimization algorithms, including gradient descent [19] and simulated annealing [20,21], with on-demand customization and interactive manipulations. Through real-time rendering of a live interaction map with smooth color ramping to capture the distance matrix alongside its compatible 3D conformation color-coded to highlight the secondary structural geometry, PolyFold makes it possible to dynamically view the folding process. Additionally, an interactive movement panel provides the ability to structurally manipulate the molecule. PolyFold does not require familiarity with protein biochemistry and provides an easily accessible platform for elucidating the distance-based protein folding process.

## PolyFold Features

As shown in **Figure 1**, the PolyFold GUI consists of three main panels: a live interaction map panel for visualizing the target distance matrix (upper triangle) and the distance matrix currently realized (lower triangle) with real-time updates, a dynamic structural display panel rendering the 3D conformation of the protein molecule compatible with the current distance matrix, and a movement panel that permits users to interactively manipulate the molecule. The core of PolyFold is implemented in Java, and the GUI controls make extensive use of the JavaFX application library. The code is cross-platform and builds and runs on macOS, Linux, and Windows. Pre-packaged binaries are also available for plug-and-play execution (see section 2 in **S1 Text**).

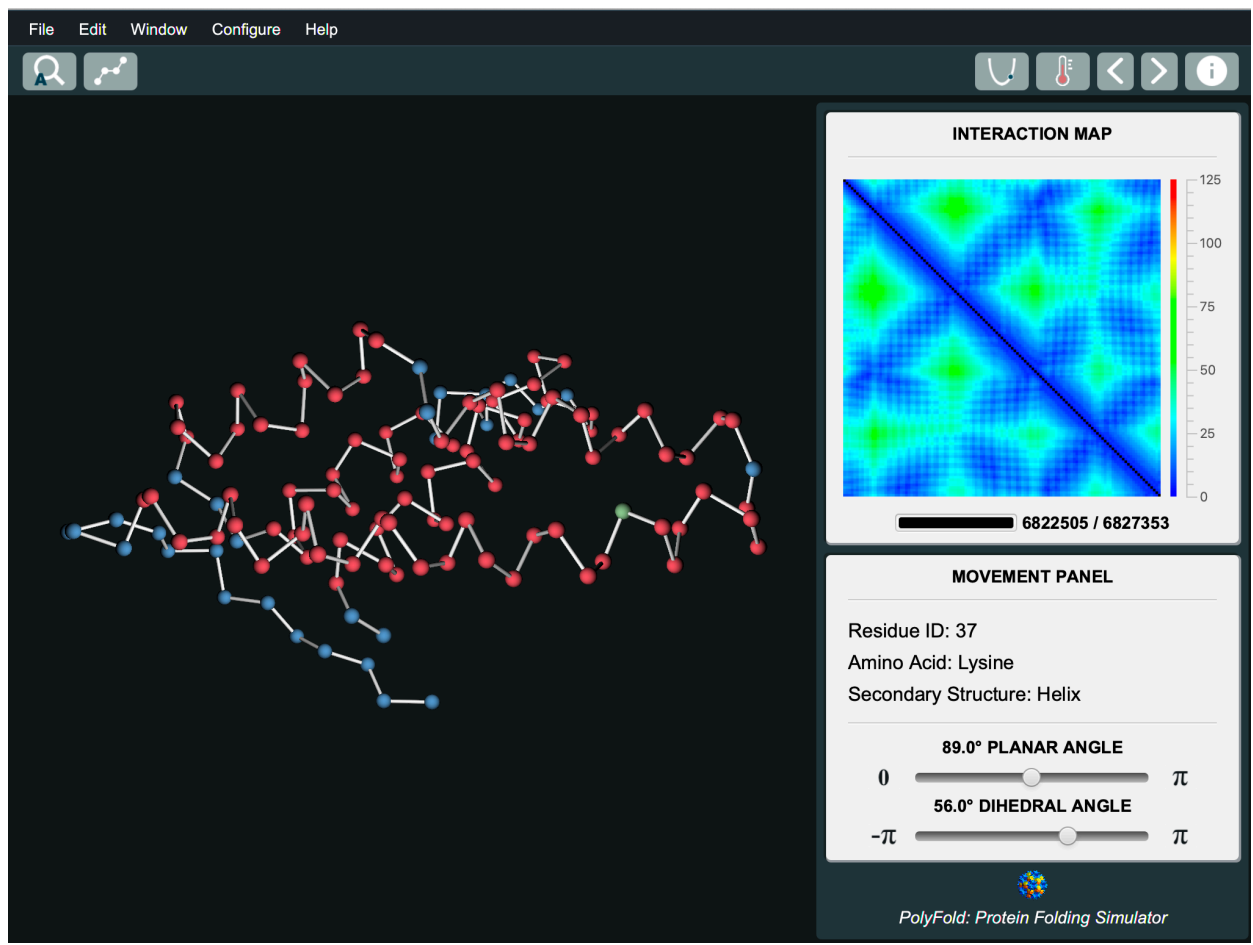

**Figure 1.** A representative PolyFold distance-based folding session for the amino terminal domain of enzyme I from escherichia coli (PDB ID: 1zym), with real-time display of the interaction map and its compatible 3D structure. Residue 37 is selected for manipulation.

PolyFold includes two optimizers for distance-based folding: gradient descent and simulated annealing, the former operating in Cartesian space and the latter in angular space. Users can launch interactive versions of both optimizers, which dynamically update the display as they run and can be cancelled prior to completion. Cascaded runs which either alternate or repeat optimizers, such as the repeated gradient descent used in AlphaFold [7], are also possible. The parameters of both optimizers are fully configurable (see section 3.3 in **S1 Text**). It is worth mentioning that PolyFold's optimization engine is designed for real-time and interactive visualization of the distance-based optimization as opposed to physics-based Molecular Dynamics (MD) simulations, which are often used for intermediate state or pathway analysis.

PolyFold's custom interactive manipulations have been specifically implemented for real-time manipulation of a molecule. Users are able to manipulate the molecular geometry in real-time with a simultaneous update to the live interaction map by selecting a residue and updating its pseudo planar and dihedral angles by dragging sliders. This feature can be particularly useful for multi-domain proteins by folding the full-length structure via distance-based optimization and subsequently adjusting the relative domain orientations by manipulating the domain linkers. PolyFold also keeps track of a history of modified states for undoing, redoing, saving, loading, and restoring the molecule to an unfolded extended state during various stages of interactive manipulations or in-built optimizations. Structures can be translated, rotated, scaled, and auto-zoomed as needed (see section 3.1 and 3.2 in **S1 Text**).

A PolyFold session can be started by supplying a distance matrix similar to the biannual Critical Assessment of protein Structure Prediction (CASP) [22–25] experiments' residue-residue (RR) format along with secondary structures (see section 4.1 in **S1 Text**). Intermediate session states can be saved to anonymous save states for quick recall or saved to named save states for lengthier sessions. Further, structures can be saved in Protein Data Bank (PDB) [26] format and restored in a later session. Prior to saving a structure, PolyFold performs secondary structure-assisted geometric chirality checking using a heuristic cost function for identifying the correct chirality as the sum over tetrapeptides in α-helices and β-sheets [27]. While PolyFold can work for large proteins, reasonably sized structures (length < 500 residues) are currently supported for seamless rendering.

## Case Study

To examine the accuracy and robustness of PolyFold for distance-based folding, we study the folding of a Ribosomal protein 1ctfA of length 68 residues [28] from near-native distance matrices as well as noisy distance matrices. In all cases, we run PolyFold by employing a single run of simulated annealing with a random seed of 0 followed by three repeated runs of gradient descent with PolyFold's default parameters.

We first perform distance-based folding by feeding a near-native distance matrix into PolyFold in CASP RR format after computing the floors and ceilings of the true inter-residue distances of the target protein 1ctfA. That is, the near-native distance matrix supplied to PolyFold specifies the distances to be within 1Å of the true real-valued distances, thus simulating a reconstruction scenario. As shown in **Figure 2**, PolyFold successfully reconstructs the structure of the target protein with a very high TM-score [29] of 0.92, demonstrating the effectiveness of PolyFold's in-built optimizers.

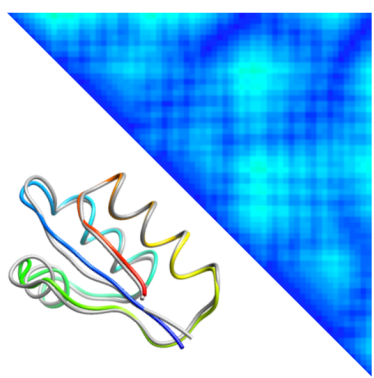

TM-score = 0.92

**Figure 2.** PolyFold distance-based reconstruction for the target 1ctfA with a near-native distance matrix. The upper diagonal shows the inter-residue distance matrix, and the lower diagonal shows the structural superimposition between the PolyFold predicted model (in rainbow) and the experimental structure of the target (in gray).

Next, we investigate the effect of feeding noisy distance matrices [30,31] into PolyFold's structural optimization engine by systematically introducing zero-mean Gaussian noise having standard deviations of $\sigma = 1, 2$, and 4Å into the true distance matrix of the target protein 1ctfA. This is accomplished by first calculating the inter-residue pair-wise distances from the near-native PDB file. We then create a pool of residue pairs (i, j) where $| i - j | > 6$. Next, we either uniform randomly select 50% of pairs or select 100% of pairs from the pool to be modified. We refer to this selection percentage as the "noise level." The calculated near-native distances are then modified with zero-mean Gaussian noise with the specified standard deviation. As shown in **Figure 3**, we observe that PolyFold's optimizations are fairly noise-tolerant, predicting the correct fold with a TM score > 0.5 [32] in all cases except in the most extreme case with noise level 100% and a standard deviation of 4Å. When noisy distance matrices with σ = 1Å are fed into PolyFold, it achieves TM-scores of 0.89 and 0.86 for 50% and 100% noise levels, respectively. By doubling the noise to σ = 2Å, PolyFold still predicts correct folds with TM-scores ≥ 0.6 for both 50% and 100% noise levels. Finally, PolyFold predicts the correct fold with a TM-score of 0.5 using a quite noisy distance matrix with σ = 4Å and a noise level of 50%, demonstrating its robustness in distance-based folding when using noisy distance matrices.

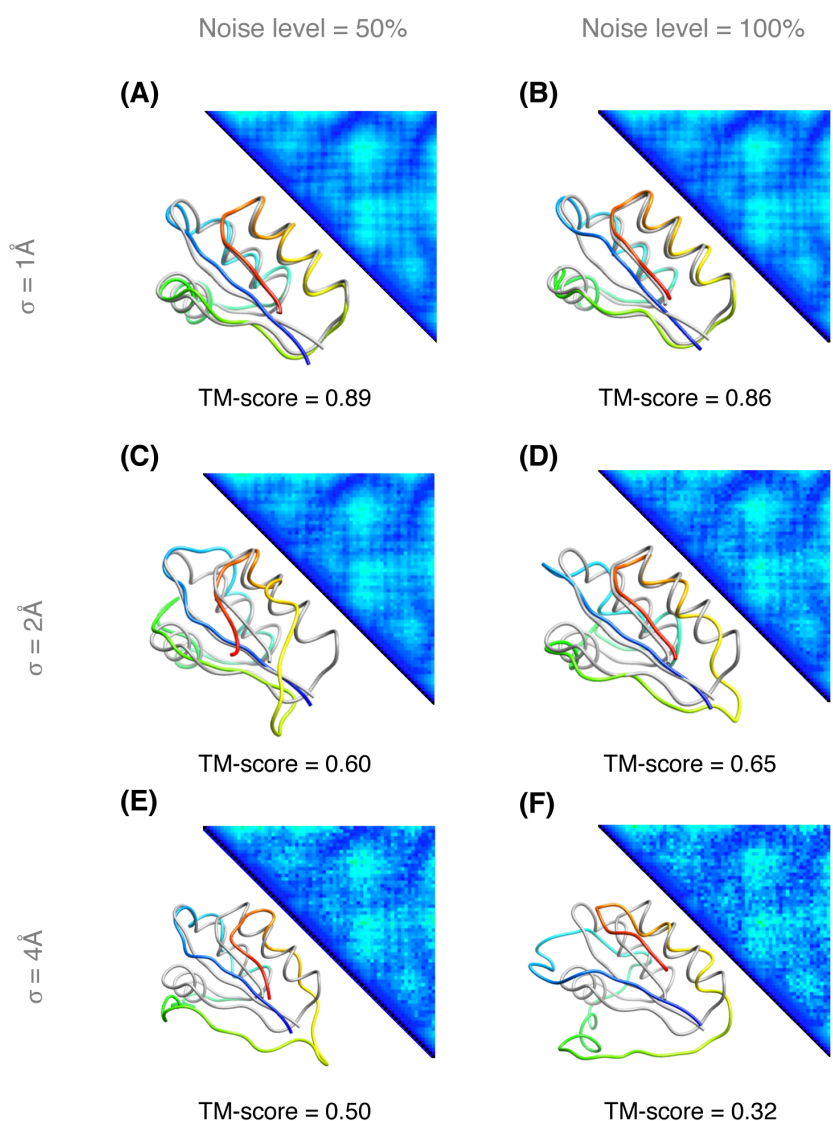
9

**Figure 3.** PolyFold distance-based folding for the target 1ctfA with noisy distance matrices. The upper diagonal shows the noisy inter-residue distance matrix by introducing zero-mean Gaussian noise into the true distance matrix with various standard deviations (σ) and noise levels. The lower diagonal shows the structural superimposition between the PolyFold predicted model (in rainbow) and the native structure of the target (in gray). (A) Noise level of 50% and σ of 1Å, (B) Noise level of 100% and σ of 1Å. (C) Noise level of 50% and σ of 2Å, (D) Noise level of 100% and σ of 2Å. (E) Noise level of 50% and σ of 4Å, (F) Noise level of 100% and σ of 4Å.

## Benchmarking

While PolyFold is primarily an interactive visual simulator for distance-based protein folding as opposed to a structure prediction method, we assess PolyFold's predictive modeling performance using a benchmark set of six small proteins ranging in length from 43 to 76 residues that have been the subject of previous studies [33,34]. For predictive modeling using PolyFold, we predict secondary structures by running SPIDER3 [35] locally. We then feed the predicted secondary structures together with distance matrices to PolyFold at varied resolutions ranging from near-native to noisy and predicted maps. In all cases, we employ two cascaded runs of PolyFold's gradient descent optimization, both for 65,000 iterations, with the first run using a step size of 0.005 and the second run using a step size of 0.0001. The optimized structural models are subsequently saved in PDB format for assessment. First, we feed near-native distance matrices within 1Å of the true real-valued distances along with predicted secondary structures into PolyFold and evaluate the accuracy of the predicted models. Next, we assess the predictive performance of PolyFold using noisy distance matrices. We follow the same strategy of introducing zero-mean Gaussian noise as discussed before to generate noisy distance matrices for the benchmark set. We feed the noisy distances matrices having standard deviations of σ = 1, 2, and 4Å at 50% and 100% noise levels together with the predicted secondary structures into PolyFold and evaluate the folding performance. Finally, we investigate the predictive ability of PolyFold when predicted distance matrices and predicted secondary structures are supplied as input. For each protein target in the benchmark set, we predict inter-residue distance maps by feeding the multiple sequence alignments (MSA) [36] of the target proteins into trRosetta [10] and then supply the predicted distances maps together with the predicted secondary structures into PolyFold to evaluate the accuracies of the predicted models. trRosetta [10] is a state-of-the-art deep learning-based protein structure prediction method that predicts inter-residue distances and orientation (dihedral and planer angles), which are subsequently transformed into restraints to generate 3D structures using energy minimization. From the standpoint of folding, both trRosetta and PolyFold rely on gradient-based optimization. However, trRosetta-based folding utilizes both distance and orientation information, whereas PolyFold uses only distance information. For a fair performance comparison with PolyFold, we, therefore, employ trRosetta-based folding using only predicted distance maps but no orientation information. We run trRosetta-based distance-only folding locally by setting the parameter ('--no-orient') that uses the same trRosetta-predicted distance maps supplied to PolyFold, albeit without orientation. Additionally, we compare the predictive modeling performance of PolyFold with two state-of-the-art protein structure prediction pipelines: I-TASSER [37,38] and Robetta [39]. We submit jobs to the I-TASSER server (https://zhanglab.ccmb.med.umich.edu/I-TASSER/) after excluding homologous templates with 30% sequence identity cutoff with the target protein and collect the top predicted model for each target protein. We submit jobs to the Robetta structure prediction server (https://robetta.bakerlab.org/) by selecting the 'AB only' option to use the Rosetta fragment assembly method for *ab initio* folding [16] and collect the top predicted model for each target protein. Of note, unlike the head-to-head comparison between the distance-based folding using PolyFold and trRosetta, a direct comparison between PolyFold and I-TASSER or Robetta is not fair because I-TASSER and Robetta have clear advantages in their use of template and/or fragment information as well as other structural features such as solvent accessibility. Furthermore, both I-TASSER and Robetta servers employ a full-fledged structure prediction pipeline by performing time-consuming conformational sampling to generate a large pool of structural decoys followed by optimal decoy selection and all-atom refinement. By contrast, PolyFold does not have such advantages since it employs computationally inexpensive distance matrix optimization over a single session while operating on a singular structure without having access to other structural features such as templates or fragments and does not perform all-atom refinement. Nonetheless, the comparison between PolyFold and I-TASSER or Robetta offers some interesting insights.

**Table 1** reports the predictive modeling performance of PolyFold using distance matrices at varied resolutions compared to I-TASSER and Robetta, as well as a head-to-head comparison between PolyFold and distance-only trRosetta, both using the same predicted distance matrices. Using predicted secondary structures and near-native distance matrices, PolyFold attains a mean TM-score of 0.73, which is higher than both I-TASSER and Robetta having mean TM-scores of 0.72 and 0.67, respectively. Moreover, PolyFold's accuracy range (maximum TM-score of 0.96, minimum TM-score of 0. 49) is better than that of I-TASSER (maximum TM-score 0.9, minimum TM-score 0.42) , and Robetta (maximum TM-score 0.82, minimum TM-score 0.42.

That is, PolyFold-based predictive modeling using near-native distance matrices delivers better performance than I-TASSER and Robetta. When noisy distance matrices ($\sigma=1$Å, noise level = 50%) are fed into PolyFold, the mean TM-score becomes 0.67, the same as the mean TM-score of Robetta. As we increase $\sigma$ and noise levels of the input distance matrices, the mean TM-scores steadily decrease. This is expected, and it demonstrates the robustness of the PolyFold's optimization engine. When predicted secondary structures and predicted distance matrices are fed into PolyFold, it attains a mean TM-score of 0.39, which is better than the distance-only trRosetta having a mean TM-score of 0.35. The better average performance of PolyFold compared to distance-only trRosetta underscores the effectiveness of PolyFold's gradient-based optimization. Interestingly, predicted distance matrices lead to better average accuracy of the resulting structural models in PolyFold than the noisy distance matrices with $\sigma=4$Å with noise level 50% and 100%; whereas the use of noisy distance matrices with $\sigma=2$Å results in an average TM score > 0.5 for both 50% and 100% noise levels, thus outperforming modeling with predicted distance matrices. That is, the quality of the predicted distance matrices possibly lies somewhere in between the qualities of the noisy distance matrices at $\sigma=2$Å and $\sigma=4$Å. The results indicate that PolyFold's optimization engine is sensitive to subtle changes in the quality of the input distance matrix and therefore may be suitable for studying the impact of noisy and predicted distance matrices in protein modeling to investigate which parts of the protein are over or under-restrained. These regions of the distance matrix could then be modified as appropriate in order to improve predictive modeling performance. Furthermore, PolyFold's fully customizable optimization engine enables users to experiment how various optimization parameters such as the step size of gradient descent might affect the resulting structural models in real time. This may help with modeling flexible regions such as loops that may be under-restrained in a predicted distance matrix. In summary, PolyFold is robust, versatile, and practically useful for predictive protein modeling.

**Table 1.** Predictive modeling performance on the benchmark dataset using PolyFold with SPIDER3 predicted secondary structures and near-native, noisy, and predicted maps. I-TASSER and Robetta *ab-initio* modeling results, obtained by submitting jobs directly to their web servers, as well as distance-only trRosetta results, obtained by running it locally with parameter settings ('--no-orient'), are also reported. In all cases, the mean, maximum and minimum TM-scores of the top predicted models are reported. Values in bold represents the best performance.

| Methods | Mean | Maximum | Minimum |
| --- | --- | --- | --- |
| PolyFold w/ near-native maps | **0.73** | **0.96** | **0.49** |
| I-TASSER | 0.72 | 0.9 | 0.42 |
| Robetta *ab-initio* | 0.67 | 0.82 | 0.42 |
| PolyFold w/ noisy maps ($\sigma$=1Å, noise level=50%) | 0.67 | 0.87 | 0.49 |
| PolyFold w/ noisy maps ($\sigma$=1Å, noise level=100%) | 0.66 | 0.87 | 0.42 |
| PolyFold w/ noisy maps ($\sigma$=2Å, noise level=50%) | 0.56 | 0.78 | 0.41 |
| PolyFold w/ noisy maps ($\sigma$=2Å, noise level=100%) | 0.55 | 0.72 | 0.32 |
| PolyFold w/ noisy maps ($\sigma$=4Å, noise level=50%) | 0.33 | 0.44 | 0.24 |
| PolyFold w/ noisy maps ($\sigma$=4Å, noise level=100%) | 0.26 | 0.31 | 0.2 |
| PolyFold w/ predicted maps | 0.39 | 0.49 | 0.3 |
| trRosetta (distance-only) | 0.35 | 0.63 | 0.27 |

## Conclusions

PolyFold offers a real-time visual simulator for capturing the optimization processes of distance-based protein folding in a dynamic and interactive interface. Being robust and resilient to noise in distance matrices, PolyFold provides a versatile platform for visualizing distance-based protein folding. In the future, PolyFold may be extended to incorporate more features into the GUI for improved user experience such as multi-directional rotations of the structure or to interactively manipulate and possibly de-noise predicted distance matrices. In conclusion, PolyFold's fully configurable, robust structural optimization and manipulation engine coupled with its easy-to-use intuitive graphical interface make it accessible to both researchers and non-experts, enabling scientists to gain new insights into protein folding and facilitating broader participation.

## Acknowledgement

The authors thank the middle and high school students as well as their teachers and parents or guardians participating in the Auburn Engineering Day ("E-day") for using PolyFold and providing feedback on its interface and features.